\documentclass[letterpaper]{article} % DO NOT CHANGE THIS
\usepackage{aaai23}  % DO NOT CHANGE THIS
\usepackage{times}  % DO NOT CHANGE THIS
\usepackage{helvet}  % DO NOT CHANGE THIS
\usepackage{courier}  % DO NOT CHANGE THIS
\usepackage[hyphens]{url}  % DO NOT CHANGE THIS
\usepackage{graphicx} % DO NOT CHANGE THIS
\usepackage{natbib}  % DO NOT CHANGE THIS AND DO NOT ADD ANY OPTIONS TO IT
\usepackage{caption} % DO NOT CHANGE THIS AND DO NOT ADD ANY OPTIONS TO IT
\usepackage{algorithm}
\usepackage{algorithmic}
\usepackage{xcolor}

\usepackage{newfloat}
\usepackage{listings}
\DeclareCaptionStyle{ruled}{labelfont=normalfont,labelsep=colon,strut=off} % DO NOT CHANGE THIS
\floatstyle{ruled}
\newfloat{listing}{tb}{lst}{}
\floatname{listing}{Listing}
\usepackage{enumitem}

\title{\emph{Maestro}: A Gamified Platform for Teaching AI Robustness}

\author{
    Margarita Geleta, %\textsuperscript{\rm 1}, 
    Jiacen Xu, %\textsuperscript{\rm 1}, 
    Manikanta Loya, %\textsuperscript{\rm 1},
    Junlin Wang, %\textsuperscript{\rm 1},
    Sameer Singh, %\textsuperscript{\rm 1}, 
    Zhou Li, %\textsuperscript{\rm 1}, 
    Sergio Gago-Masague%\textsuperscript{\rm 1}
}
\affiliations{
    University of California, Irvine\\ Irvine, California 92697, USA\\
    \{mgeleta, jiacenx,manikanl,junliw1,sameer,zhou.li,sgagomas\}@uci.edu

}

\begin{document}

\maketitle

\begin{abstract}
Although the prevention of AI vulnerabilities is critical to preserve the safety and privacy of users and businesses, educational tools for \emph{robust AI} are still underdeveloped worldwide. We present the design, implementation, and assessment of \emph{Maestro}. \emph{Maestro} is an effective open-source game-based platform that contributes to the advancement of \emph{robust AI} education. \emph{Maestro} provides \emph{goal-based scenarios} where college students are exposed to challenging life-inspired assignments in a \emph{competitive programming} environment. We assessed \emph{Maestro}'s influence on students' engagement, motivation, and learning success in \emph{robust AI}. This work also provides insights into the design features of online learning tools that promote active learning opportunities in the \emph{robust AI} domain. We analyzed the reflection responses (measured with Likert scales) of 147 undergraduate students using \emph{Maestro} in two quarterly college courses in AI. According to the results, students who felt the acquisition of new skills in robust AI tended to appreciate highly \emph{Maestro} and scored highly on material consolidation, curiosity, and maestry in \emph{robust AI}. Moreover, the leaderboard, our key gamification element in \emph{Maestro}, has effectively contributed to students' engagement and learning. Results also indicate that \emph{Maestro} can be effectively adapted to any course length and depth without losing its educational quality.
\end{abstract}

\section{Introduction}
AI is becoming the key technological asset for fueling automated decision-making systems, with significant societal impact and consequences upon many fields, including such sensitive ones as medicine, transportation and education \cite{JRC_EU}. Traditionally, these systems have been designed and developed with the assumption that the environment is benign during both, their training and evaluation. However, a real environment can take adversarial shades. Consequential vulnerabilities in the shape of adversarial attacks may jeopardize AI-based systems and lead to unexpected outputs. Adversarial attacks have been extensively developed in computer vision applications and consist in altering input data to increase the odds of misclassification by the model \cite{goodfellow2017attacking, adv_attacks, adv_attacks2, net_properties, adv_attacks3}. Potential vulnerabilities in fast-spreading AI-based systems, likewise in the cybersecurity field, have created concerns and motivated new directions and strategies  to educate the future AI workforce in countermeasures and the prevention of AI vulnerabilities. This aforementioned domain of expertise is known as \emph{robust AI}. Despite \emph{robust AI} being an active area of research, the advancement of educational tools specifically targeting this domain is still in its infancy. In fact, we found there were attempts to teach AI in general by reproducing the existing algorithms~\cite{lucic2022reproducibility}, using explicit ethical agent~\cite{green2021ai}, playing value cards~\cite{shen2021value}, programming robots~\cite{mataric2007materials}, etc., but none of them touches the area of \emph{robust AI}, not to mention building a learning platform for \emph{robust AI}\footnote{We found there are academic and industrial platforms to test and evaluate \emph{robust AI} algorithms, like CleverHans (\url{https://github.com/cleverhans-lab/cleverhans}) and ART (\url{https://github.com/Trusted-AI/adversarial-robustness-toolbox}), but none of them are organized under \emph{goal-based scenarios} (GBSs), as explained later.}. Hence, we make the \emph{first} step towards bridging this gap. To facilitate other educators to teach \emph{robust AI}, we will release the platform publicly in a GitHub repository.

There exist many examples of learning platforms and containerized environments in the cybersecurity domain, but very few in AI. Previously developed educational tools targeted to explore and simulate cybersecurity flaws and develop secure coding practices to train new cybersecurity professionals \cite{kjorveziroski2020cybersecurity, cyber_learning, acosta2017platform, kalyanam2017try, mirkovic2012teaching}. A variety of open-source educational tools to teach cybersecurity have been confidently introduced into secondary and higher education around the world, such as the SEED project \cite{du}  or the OWASP Juice Shop project \footnote{OWASP Juice Shop (\url{https://owasp.org/www-project-juice-shop/}) is an open-source project for demonstration and exercise in security risks in modern web applications.}. The common denominator of all these educational projects is the principle of \emph{learning by doing}, i.e., providing a hands-on experience where users are presented to \emph{goal-based scenarios} (GBSs) \cite{gbs, reigeluth2013instructional}.

GBSs provide learning in a contextualized fashion, since they marry knowledge and skills with real-life scenarios, supplying a more effective way of knowledge consolidation and skill development. Educational projects following GBSs optimize over the intended skills to learn by exposing students to opportunities where they can practise those in a task environment \cite{gbs}. Essentially, students become active participants in the learning scenario \cite{watson2011case}, which stimulates them to move towards the completion of the specified task. This encouragement, also referred to as \emph{motivation}, is the key ingredient in pedagogical success \cite{uguroglu1979motivation, reigeluth2013instructional, deci2013intrinsic, TOHIDI2012820, ahmet2013inspiring}. Games and competitions, as educational tools, have been shown to encourage curiosity and motivation for learning \cite{BURGUILLO2010566, gamification_in_education, lepper1988motivational, watson2011case}, and have been actively used in schools, colleges and universities for computer science curricula \cite{10.1145/274790.273140, becker, ebner2007successful, data_structs}. In particular, in courses requiring programming, the concept of \emph{competitive programming} has been developed \cite{data_structs}. \emph{Competitive programming} allows students to engage with the programming topics set by the instructor. By improving and evaluating students' skills in a coding tournament, students can either try to beat a baseline set by the instructor or compete against other students. Examples of \emph{competitive programming} implementations and communities include \emph{Jutge.org} \cite{jutge}, % used in university teaching and olympiads in informatics; 
\emph{CodeWars} \cite{codewars}, \emph{CodeChef} \cite{codechef}, \emph{SPOJ} \cite{spoj}, \emph{LeetCode} \cite{leetcode}, among others. \emph{Competitive programming} has also been adapted in data science and machine learning contexts, with competitions where students compete not for a compile/no compile, best time execution and memory usage, but compete for attaining better supervised learning performance metrics, such as accuracy. Examples of such competition platforms include \emph{Kaggle} \cite{kaggle} and \emph{Driven Data} \cite{drivendata}.

\section{Research Goals}
The present project goal is to design, implement and evaluate an innovative open-source educational tool (\emph{Maestro}) for training courses in \emph{robust AI}. The goal of \emph{Maestro} is to provide a friendly, \emph{competitive programming} environment to engage undergraduate and graduate students in course assignments to acquire skills in \emph{robust AI}. In our work, we investigate (i) whether \emph{Maestro} provided an efficient learning platform for students to interact and acquire skills in \emph{robust AI}, and (ii) whether \emph{Maestro} provided students with enough autonomy for training on \emph{robust AI} as a self-paced learning tool.

\section{Approach}
 To encourage active student participation, we propose using game-based assignments \cite{gamification_in_education}, which include game stories, use cases and standard scenarios to reproduce settings covered by \emph{Maestro}. The game is structured as an engaging project competition, where each student or team is assigned roles of attacker or defender, and \emph{Maestro} assigns scores and ranks participants based on the quality of their attack or defense. These results are displayed in a user-friendly interface. Specifically, in the course described in this study, students are introduced to three different GBSs: one for adversarial attacks referred to as \emph{Attack phase}; another one for adversarial defenses -- \emph{Defense phase}, and lastly an interactive scenario, referred to as \emph{War phase}, where attacker and defender teams compete among each other, following the \emph{Build it, break it, fix it} strategy \cite{parker2020build}.

\subsection{Design}
In order to maximize the instructional benefits of \emph{Maestro}, it has been designed following the intrinsic motivational principles \cite{lepper1988motivational, xerox_fun}. The basic proposition of these principles lies in the fact that interest and motivation can be stimulated by challenge in a contextualized environment \cite{deci2013intrinsic}. A continuously challenging activity can be given if (a) the student is uncertain about their success outcome and if (b) the student receives feedback about their progress towards the task goal. Within \emph{Maestro}, students are enrolled into a \emph{competitive programming} environment where they need to beat the baselines (i.e., hidden models) and their fellow students' solutions, contributing to the challenge of the game. Students also have access to the \emph{Maestro} leaderboard, where their scores, ranks and/or (possible) errors are displayed, providing feedback and exposing the (possible) inconsistencies in their submitted solutions.

\subsubsection{Interaction diagram}
\begin{figure}[!b]
 \centering
     \includegraphics[width=0.85\linewidth]{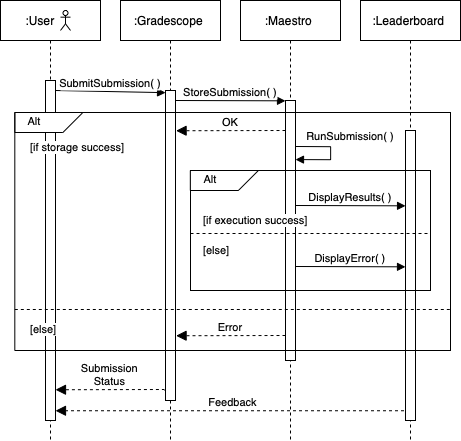}
     \caption{Interaction diagram of \emph{Maestro}.}
     \label{fig:interaction-diagram}
\end{figure}
The storyboard sequence diagram (Figure \ref{fig:interaction-diagram}) represents the lifeline of \emph{Maestro} usage. A student, once they developed their attack or defense according to the task defined in the GBS, submits their own solution to Gradescope\footnote{Gradescope (\url{https://www.gradescope.com}) is a suite of tools designed to accommodate the grading workflow.}. Gradescope is a user-friendly platform with a submission recording system that documents each submission's timestamps, files' content, and user information. \emph{Maestro} has been integrated with Gradescope in order to store the students' codes and provide an easy access to them for both, students and instructors, as well as join students into teams. After a solution has been sent to Gradescope, if it has been successfully stored, the autograder of Gradescope forwards the student's submission to the \emph{Maestro} server.
\emph{Maestro} executes the submission and benchmarks the provided solution against several hidden models. Once the \emph{Maestro} server finishes the evaluation, a comma-separated values (CSV) file with the results of these benchmarks is generated. That CSV file is displayed on the \emph{Maestro} leaderboard for the corresponding phase (\emph{Attack}, \emph{Defense} or \emph{War}), which students can access anytime. In case that the submission errors during runtime at \emph{Maestro}, the caught exception is displayed at the error boards on the \emph{Maestro} leaderboard.

\subsubsection{System diagram}

\begin{figure}[!b]
 \centering
     \includegraphics[width=\linewidth]{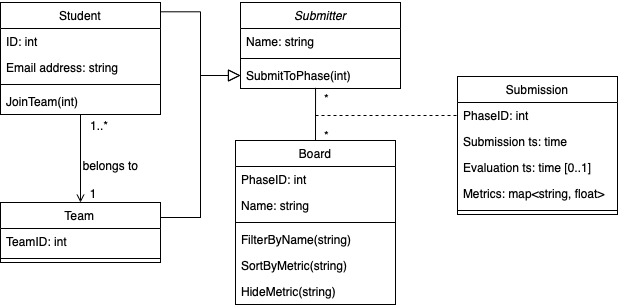}
     \caption{\emph{Maestro} Unified Modeling Language (UML) entities.}
     \label{fig:uml}
\end{figure}
Because a submission can be done by an individual student or a group of students, each submission is linked to a generalized entity referred to as \emph{submitter} (Figure \ref{fig:uml}). Each \emph{submitter} can have an unlimited amount of submissions for each phase prior to the submission deadline, and each phase has an associated \emph{Maestro} leaderboard. Once a submission has been stored and sent to \emph{Maestro} (see Figure \ref{fig:interaction-diagram}), the \emph{submitter}'s code is executed and tested against several hidden models. Depending on the phase, if the submission is for an attack, \emph{Maestro} tests its effectiveness to downgrade the classification accuracy of hidden vision models, such as the convolutional neural network LeNet~\cite{lecun1998gradient}. 
In the case that the submission is for a defense, such as adversarial training~\cite{goodfellow2014explaining}, \emph{Maestro} tests its robustness against several white-box or black-box adversarial attacks, namely FGSM~\cite{goodfellow2014explaining} and PGD~\cite{madry2017towards}. Finally, if the submission is for an attack or defense in the war phase, the provided solution is tested against the best attack or defense submissions from previous phases. At the end, \emph{Maestro} associates a map with evaluation metrics to each submission. By default, the boards display only the last evaluated submission by each \emph{submitter}. However, each board implements a set of operations, which includes filtering by \emph{submitter}'s ID. This allows submitters to see their submission history and analyze the provided feedback to implement strategies to improve their following submission.

\subsubsection{Methods}
The platform consists of code templates and essential libraries to let the students implement solutions for the attack homework (e.g., Genetic Algorithm \cite{chen2019poba}),
% Fast Gradient Sign Method \cite{fastgradient}),
defense homework (e.g., Adversarial Training \cite{goodfellow2014explaining}), attack project (where the student can implement any attack method against a selected defense method), defense project (where the student can implement any defense method against a selected attack method) and war phase (where the student plays attack and defense in turns). In further depth, the assignments are built under different scenarios such as evaluating the robustness of the rainforest monitor systems. Adversarial methods are used in the context of multi-class classification, using MNIST \cite{LeCun2005TheMD} and CIFAR-10 \cite{Krizhevsky2009LearningML} datasets, respectively.

\subsubsection{Metrics}
\begin{figure}[!]
 \centering
     \includegraphics[width=\linewidth]{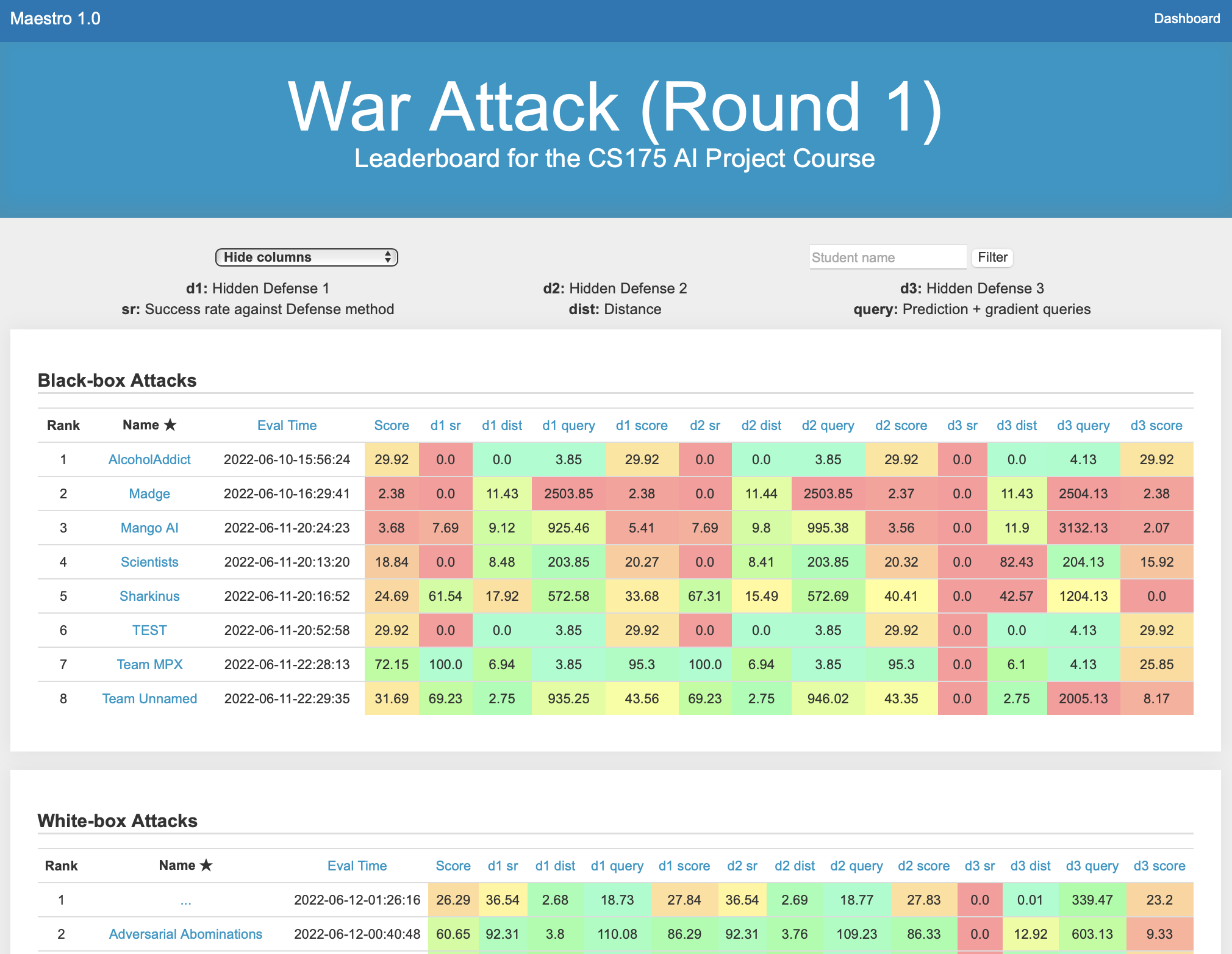}
     \caption{\emph{Maestro} leaderboard showing the \emph{War} phase board. }
     \label{fig:leaderboard}
\end{figure}
We evaluate each submission with metrics that are widely adopted by the literature of robust AI:

\begin{itemize}
    \item \textbf{Efficiency}: We consider the runtime overhead and resource consumption for the efficiency metric. For an attack submission, we encourage the implementation to trigger less interactions with the AI system (measured by the number of queries) and execution in shorter time. For a defense submission, the implementation needs to harden the AI model and we measure the extra time that is consumed for this process.
    
    \item \textbf{Effectiveness}: For an attack submission, we measure how it decreases the prediction accuracy of a targeted model, and how stealthy the attack sample is (e.g., measured by $L_2$ distance that compares the original image and the perturbed image~\cite{carlini2017towards}). For a defense submission, we consider the change of the prediction accuracy; a good implementation should keep a similar level even under attack.
    
\end{itemize}

On top of these metrics, we also compute a weighted sum as the overall score for a submission. When multiple submissions are required (e.g., during the war phase), the overall score is a weighted sum of all the submissions.

\subsubsection{Leaderboard}
To make the usage of \emph{Maestro} compelling, a leaderboard is a key feature to include in order to gamify the learning environment, as it fulfills the learners' needs of connectedness (a tool driving to interaction), competence (a tool driving to practice and mastery), and autonomy (a tool for partial control) \cite{fotaris2016climbing, ortiz2019gamification, sailer2017gamification}. Hence, the leaderboard becomes a game design element that provides a challenging environment for the students by creating social pressure to increase their level of engagement and participation. Within the \emph{Maestro} leaderboard (Figure \ref{fig:leaderboard}), we included the following features to ease the interaction with the data:
\begin{itemize}
    \item \textbf{Submission history}: by default, each phase leaderboard displays one submission per each individual \emph{submitter}. The main reason for this design decision is to keep the board not overflown with too many submissions. However, if desired, it is possible to access the submission history of each \emph{submitter} by clicking (filtering) on their name.
    \item \textbf{Default \& custom sorting by metrics}: the sorting of all the submissions by default is by evaluation time, which translates the submission history into a chronological order and allows tracking the improvement over different metrics. At anytime it is possible to sort the boards by any other metric by clicking on the metric name at the header.
    \item \textbf{Search by name}: the leaderboard includes a search input for fast retrieval of searched rows. The routine behind filters the submission data to display only the rows that match the search query.
    \item \textbf{Color indicators \& thresholds}: each leaderboard has a configuration file where minimum and maximum values for each metric can be predefined. Additionally, it is possible to define a threshold value. Cells containing numbers above the threshold will be colored in green tones (suggesting an accomplishment regarding that metric), and cells with numbers below the threshold  will be colored in red tones (suggesting further improvement of that metric). That color gradient allows for a fast visual evaluation of the submissions.
    \item \textbf{Metric selection}: leaderboards can display only the desired metrics which can be selected using a dropdown menu located at the top area of the leaderboard.
\end{itemize}

\subsection{Evaluation}
In order to assess the effectiveness of this new GBS-based approach for \emph{robust AI} education, we  implemented a course with \emph{Maestro} for undergraduate students, which allowed the evaluation of its performance based on several metrics such as learning outcomes and students' engagement. For that aim, we designed a questionnaire, surveying upon the attitude and perceptions towards the \emph{Maestro} platform as a \emph{robust AI} educational tool, using well-established measures of course engagement and psychometric scales \cite{mandernach2011assessing, handelsman2005measure}. Indicator statements and questions have a restricted amount of answers measured on a Likert scale (1--5). 

The main survey consisted of twelve questions. Each question has a prefix \textbf{Q}. Ten questions have five possible answers measured on a Likert scale ($1=$ Strongly disagree or \textbf{SD}, $2=$ Disagree or \textbf{D}, $3=$ Neutral or \textbf{N}, $4=$ Agree or \textbf{A}, $5=$ Strongly agree or \textbf{SA}). Two other questions have a binary and categorical response, respectively. To assess the quality and the perception of the leaderboard as a key gamification feature of \emph{Maestro}, we included five additional questions in our survey specifically targeting the leaderboard features. These questions have a prefix \textbf{LQ}, and they are also measured on a Likert scale (1--5). 

The full set of questions can be found in \emph{Appendix A}. We can group the survey questions into four categories (\textbf{C1}--\textbf{C4}):
\begin{enumerate}[label=\textbf{(C\arabic*)},align=left]
\item \textbf{Gamification}:
    \subitem \textbf{Competence}: this set of questions aims to check the quality of \emph{Maestro}'s game elements. \textbf{Q1} and \textbf{Q3} strive to examine the constructive effect of the leaderboards and game stories, respectively.
    \subitem \textbf{Autonomy}: \textbf{Q2} examines the fulfilment of the psychological need of autonomy in regard to decision freedom.
    \subitem \textbf{Connectedness}: \textbf{Q4} examines the fulfilment of the psychological need of social relatedness. Answers aim to score the level of sensed relevance among teammates and union towards a common goal.
\item \textbf{Educational experience}:
    \subitem \textbf{Curiosity}: \textbf{Q5} raises a question about the perceptual level of interest to learn more about \emph{robust AI} catalyzed by \emph{Maestro}.
    \subitem \textbf{Consolidation}: \textbf{Q6} examines the learning success and effectiveness by the \emph{learning by doing} experience.
    \subitem \textbf{Maestry}: \textbf{Q7} extrapolates \textbf{Q6} by surveying even further, not asking just about material consolidation but also about how students feel about transferring their new skills to other AI domains.
    \subitem \textbf{Effectiveness}: \textbf{Q8} inquires about the perception of \emph{Maestro} as an appropriate tool for communicating the course content.
    \subitem \textbf{Course organization}: \textbf{Q11} inspects the preference between the classical way of instructions (evaluation via midterms) or our GBS-based instruction with \emph{Maestro} (via contextualized assignments).
\item \textbf{Academic challenge}:
    \subitem \textbf{Material quantity}: our course taxonomy can be divided into three phases (\emph{Attack}, \emph{Defense} and \emph{War}). \textbf{Q9} questions the perceived challenge by the given amount of workload and material to be learned.
    \subitem \textbf{Time usage}: \textbf{Q10} examines the perceived amount of spent time in order to finalize the assignments.
    \subitem \textbf{Tasks}: \textbf{Q12} compares the perceived difficulty in material consolidation between adversarial attacks and defenses. 
\item \textbf{Leaderboard features}: \textbf{LQ1}, \textbf{LQ2}, \textbf{LQ3}, \textbf{LQ4} and \textbf{LQ5} explore students' satisfaction with several features that were included in the \emph{Maestro} leaderboard.
\end{enumerate}

\section{Results and Discussion}

The first version of \emph{Maestro} was successfully  released in late 2021 and put into practice by 147 undergraduate students enrolled in \emph{Projects in AI} course (with focus on \emph{robust AI}) at the University of California, Irvine (UCI). %a large public university in the U.S. 
The \emph{Maestro} leaderboard was released in Spring 2022. There have been two course offerings; one in Winter (with 70 students enrolled) and a another in Spring quarter (with 77 students enrolled) both in 2022. All students had access to: (a) the private course Canvas\footnote{Canvas (\url{https://canvaslms.com}) is a learning management system (LMS).} page (in the UCI %institutional 
intranet) including assignment descriptions and game stories,
(b) the publicly available \emph{Maestro} repository on GitHub\footnote{\emph{Maestro} server can be found at \url{https://github.com/ucinlp/maestro-public}.}, and (c) the private course Gradescope page, where students could submit their assignments solutions. Students enrolled in the Spring offering had also access to the Maestro leaderboard\footnote{\emph{Maestro} leaderboard can be found at \url{https://github.com/ucinlp/maestro-leaderboard}.}. Students in the course were instructed to clone the \emph{Maestro} repository from GitHub and practise locally before submitting their solutions to the remote \emph{Maestro} server. 
At the conclusion of the courses, students participated in the survey anonymously regarding gamification, experience and challenge categories. Students from the Spring 2022 offering additionally answered questions regarding the leaderboard features category.

\begin{figure}[!htb]
 \centering
     \includegraphics[width=\linewidth]{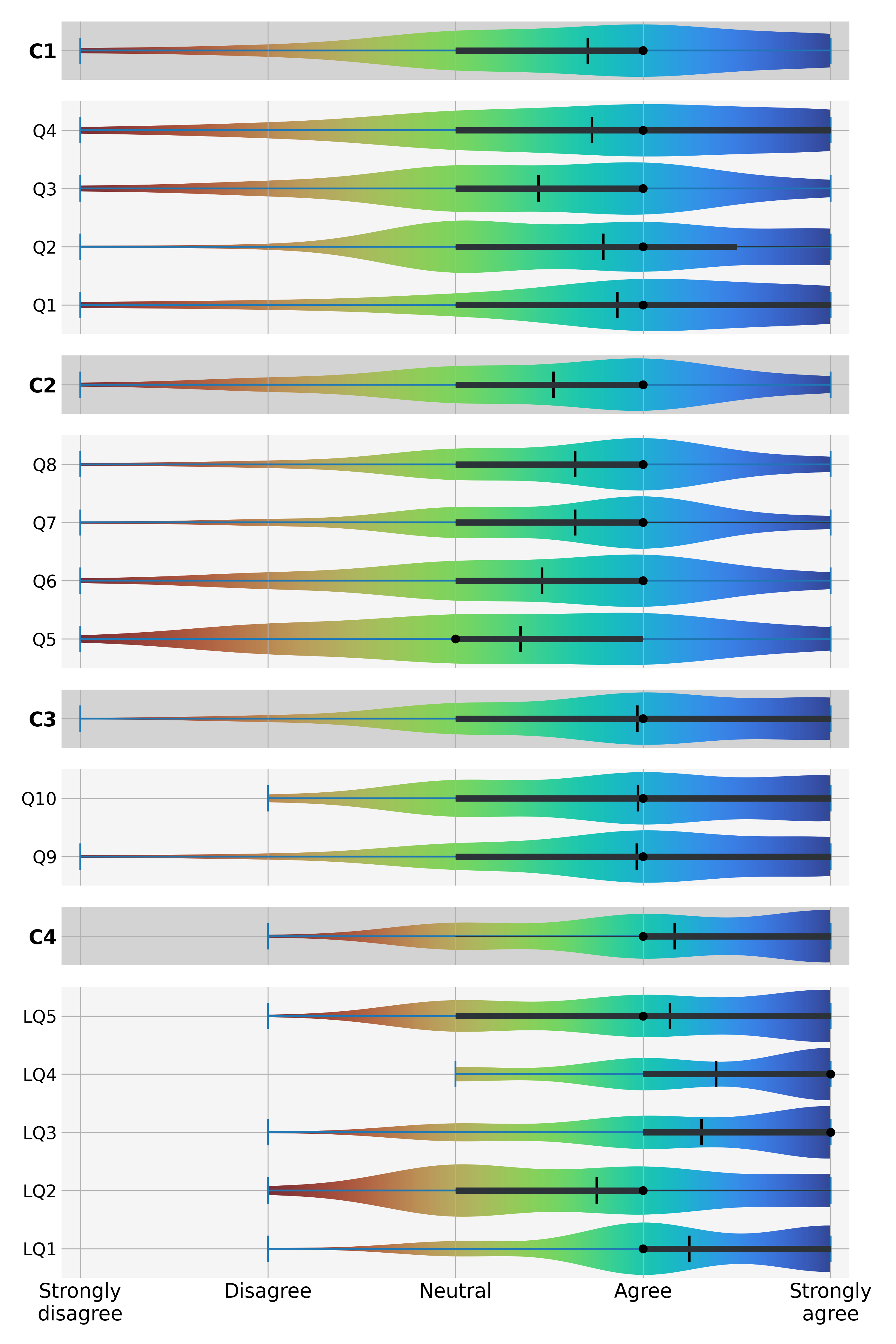}
     \caption{Violin plots featuring the responses of continuous-scale survey questions. The black rectangles in the boxplots represent the interquartile ranges, where the black dot represents the median. Black vertical lines represent the average response.}
     \label{fig:boxplots}
\end{figure}

\begin{figure*}[!htb]
 \centering
     \includegraphics[width=\linewidth]{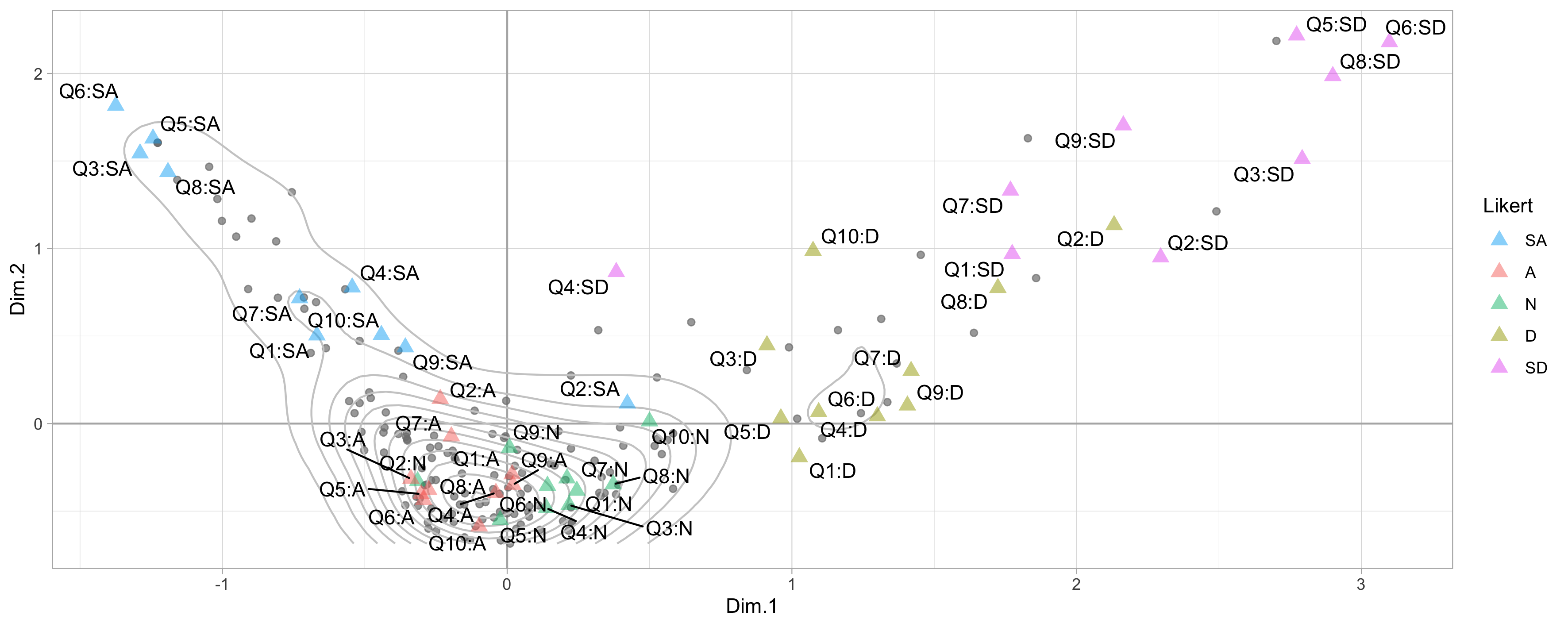}
     \caption{Multiple Correspondence Analysis (MCA) based on indicator matrix ran on the general survey results.}
     \label{fig:mca}
\end{figure*}

The results of the survey point to a significant efficacy of \emph{Maestro} as it can be observed in the violin plots from Figure \ref{fig:boxplots}, where \textbf{C1}, \textbf{C2}, \textbf{C3} and \textbf{C4} are the aggregated responses of continuous-scale survey questions from each category (\textbf{C1} $=$ Gamification, \textbf{C2} $=$ Educational experience, \textbf{C3} $=$ Academic challenge, \textbf{C4} $=$ Leaderboard features). Observe that the median of the majority of questions lies at the \emph{Agree} (\textbf{A}) answer. Regarding gamification questions, \emph{Competence} \textbf{C1} had the highest rate in positive answers, signaling the effectiveness of the leaderboards and game stories. This is reinforced by the positive answers from the second category, educational experience \textbf{C2} (especially in \textbf{Q7} and \textbf{Q8}). As far as academic challenge \textbf{C3} is concerned, there is an unanimity that \emph{Maestro} had provided a challenging learning environment. 

The leaderboard features (\textbf{LQ1}--\textbf{LQ5}) had a very positive perception, highlighting the median of \textbf{LQ3} and \textbf{LQ4} as \emph{Stronly Agree} (\textbf{SA}). According to \textbf{LQ3} and \textbf{LQ4}, the most appreaciated leaderboard features are the sorting options and the access to the submission history, while the feature of hiding columns (\textbf{LQ2}) had a perception biased towards neutral.

\begin{figure}[!htb]
 \centering
     \includegraphics[width=\linewidth]{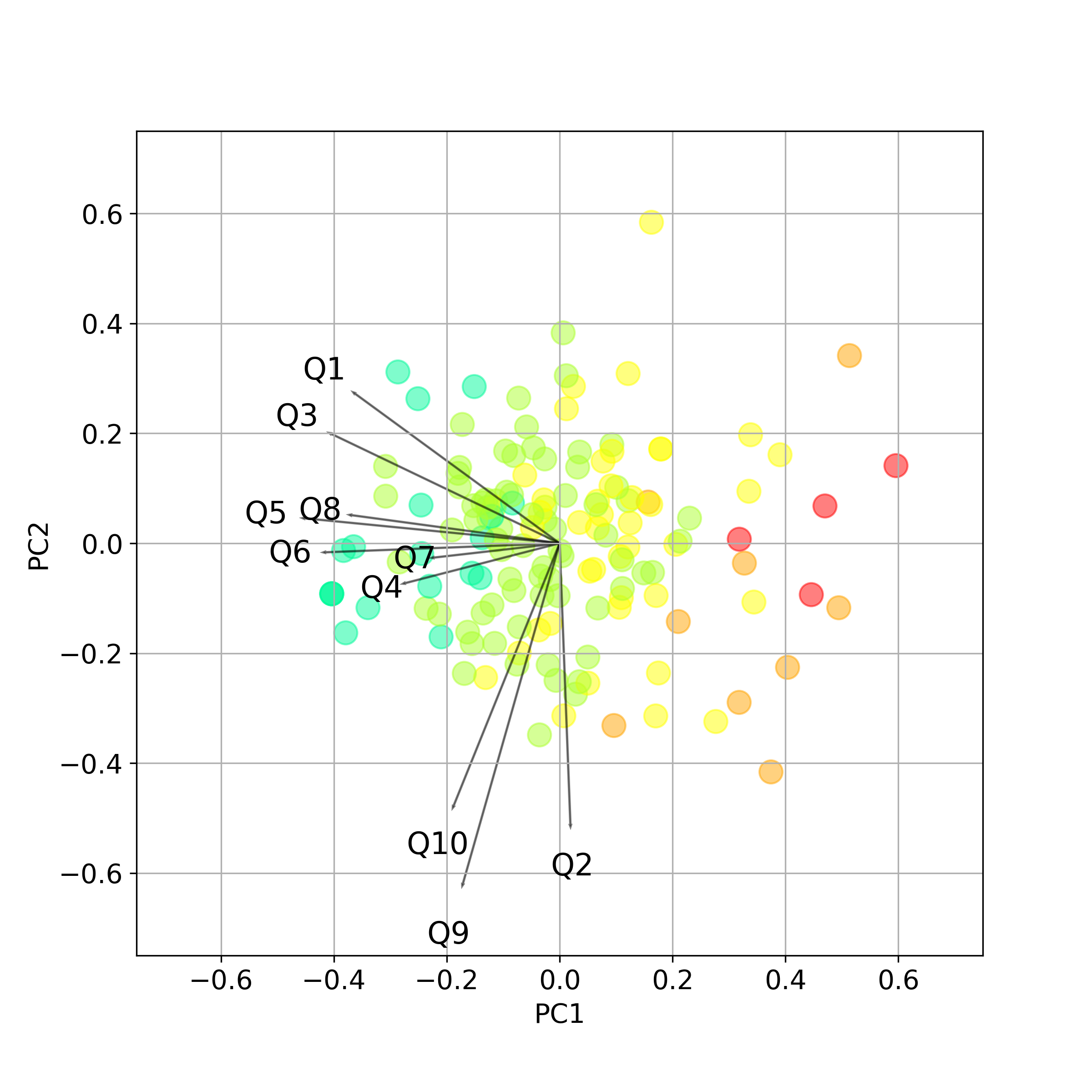}
     \caption{Biplot of Principal Component Analysis (PCA). The origin represents the answers' mean. The arrows are the directions of growth of each variable.  Dots represent each projected individual and the color gradient corresponds to the \emph{Effectiveness} variable \textbf{Q8} (with \emph{red} $ = $ Strongly disagree or \textbf{SD}, \emph{orange} $ = $ Disagree or \textbf{D}, \emph{yellow} $ = $ Neutral or \textbf{N}, \emph{green} $ = $ Agree or \textbf{A}, \emph{turquoise} $ = $ Strongly agree or \textbf{SA}).}
     \label{fig:biplot}
\end{figure}

Given the partially categorical nature of the ordinal scale of our survey answers, we opted for indicator-matrix-based Multiple Correspondence Analysis (MCA) to analyze the association between the survey answers. In Figure \ref{fig:mca} we represent questions \textbf{Q1}--\textbf{Q10} with answers measured on a Likert scale. 
For instance, \textbf{Q10:SA} represents the projection of the indicator matrix column corresponding to question \textbf{Q10} and answer \textbf{SA}. Each grey dot represents an individual that has responded the survey (147 points). Additionally, since the Likert scale can be interpreted as a continuous gradient of satisfaction, we performed Principal Component Analysis (PCA) on the survey answers, and projected both, rows and columns, in a bidimensional latent space resulting in a biplot (Figure \ref{fig:biplot}), which is a powerful tool for the graphical exploration of multivariate data.

In Figure \ref{fig:mca} we can observe 4 concentrations of categories: the blue cluster of \textbf{SA} categories, the yellow cluster of \textbf{SD} categories, the outspread sparse pink cluster of \textbf{SD} categories, and a densely-populated central cluster with \textbf{A} and \textbf{N} categories. We can affirm that there are no inversely correlated variables because there is no proximity between completely different categories and in Figure \ref{fig:biplot} there are no straight angles between variables. \emph{Curiosity} \textbf{Q5}, \emph{Consolidation} \textbf{Q6}, \emph{Maestry} \textbf{Q7} and \emph{Effectiveness} \textbf{Q8} appear to be highly correlated, which indicates that students who felt the acquisition of new skills in \emph{robust AI} tended to appreciate highly \emph{Maestro}. \emph{Competence} \textbf{Q1},\textbf{Q3} and \emph{Maestry} \textbf{Q7} appear to have a clear association (see Figure \ref{fig:mca} for \textbf{Q1}-\textbf{Q7} and Figure \ref{fig:biplot} for \textbf{Q1}-\textbf{Q3}) reinforcing the idea that game elements play a crucial role in effective learning and skill-development. Another interesting association is between \emph{Connectedness} \textbf{Q4} and \emph{Maestry} \textbf{Q7} -- those individuals who frequently collaborated with others felt more successful in learning. Naturally, those who felt challenged by the amount of material to be learned spent a significant amount of time to delve into material, which is represented by the clear correlation of \emph{Material quantity} \textbf{Q9} and \emph{Time usage} \textbf{Q10}. Coincidentally, those two variables are also noticeably correlated with \emph{Autonomy} \textbf{Q10}, signalling that those individuals who felt too much autonomy are the ones who required more time for material consolidation. In average, the perception of autonomy has been between being balanced and guided enough (Figure \ref{fig:boxplots}). \emph{Gamification} variables (\textbf{Q1},\textbf{Q3}) and \emph{Educational experience} variables (\textbf{Q5}, \textbf{Q6}, \textbf{Q7}, \textbf{Q8}) are nearly perpendicular with respect to \emph{Autonomy} \textbf{Q2} and \emph{Academic challenge} variables \textbf{Q9}, \textbf{Q10} (Figure \ref{fig:biplot}). This fact signifies that the course constraints (e.g., the sense of time usage and workload) uncorrelates with the assessment of \emph{Maestro} as an educational tool, which, as matter of course, suggests that \emph{Maestro} can be adapted to any course length and depth without losing its educational quality.

\section{Conclusion}
We successfully implemented and piloted \emph{Maestro} -- an innovative educational tool for teaching and learning AI robustness. We assessed the effectiveness of \emph{Maestro} in two offerings of an undergraduate AI course, and collected students feedback with a thoroughly designed reflection survey. Based on the results, we were able to answer our initial research questions as follows:
\begin{enumerate}[label=(\roman*),align=left]
\item \textit{Does \emph{Maestro} provide an efficient learning platform for students to interact and acquire skills in \emph{robust AI}?}

Our results suggest that \emph{Maestro} has successfully gamified the learning environment driving students to interaction, \emph{robust AI} mastery and autonomy. The game elements provided in \emph{Maestro} played a crucial role in students’ effective learning and skill-development. The \emph{Maestro} leaderboard was highly valued by students. Overall, students who felt the acquisition of new skills in \emph{robust AI} tended to appreciate highly \emph{Maestro}. Lastly, students who frequently collaborated with each other in \emph{Maestro} felt more successful in learning. 

\item \textit{Does \emph{Maestro} provide students with enough autonomy for training on \emph{robust AI} as a self-paced learning tool?}

Defining the teaching strategy on GBSs has allowed students to learn and practice their skills in an immersive experience at different paces. Students felt that autonomy has been between being balanced and enough autonomy. The results also suggest that \emph{Maestro} can be adapted to any course length and depth without losing its educational quality as the course constraints uncorrelated with the assessment of Maestro as an effective educational tool.
\end{enumerate}

\section{Future Work}
Our next step is to continue the development and assessment of \emph{Maestro} by offering it as an open-source platform to other instructors, students and the broad \emph{robust AI} community. Future users of \emph{Maestro} can test it and contribute further to its implementation and maintenance. The authors plan to launch an updated version of \emph{Maestro} during 2022-2023 academic year, and continue working on its effective integration in AI courses on and off campus.

%\section{Appendices}


\appendix

\section{Study Survey Questions}
\begin{enumerate}[label=Q\arabic*,align=left]
    \item Leaderboards and competitiveness have helped me to work harder.
    \item Do you consider that the project offered too much autonomy? ($1$\emph{ = Too guided, $2$ = Guided enough, $3$ = Balanced, $4$ = Enough autonomy, $5$ = Too much autonomy})
    \item \emph{Maestro} has helped me to reinforce the theory. 
    \item I frequently worked with other students to solve problems in class.
    \item \emph{Maestro} has developed enthusiasm and interest to learn more about course content.
    \item I was able to understand class material because I could practice in \emph{Maestro}.
    \item I feel that I developed the ability to solve real AI problems. 
    \item I consider \emph{Maestro} to be an appropriate technology tool which has been effectively used to communicate the course content. 
    \item I felt challenged by the overall amount of material to be learned? 
    \item To be successful, I have spent a significant amount of time on class material.
    \item Did you like the course organization? 
    \item Which of the following topics did you find the most difficult? 
\end{enumerate}
\begin{enumerate}[label=LQ\arabic*,align=left]
    \item I feel that the color of the scores on the leaderboards has been useful / a good feature.
    \item I feel that the feature of hiding columns using a dropdown list on the leaderboard has been useful / a good feature.
    \item I feel that the feature of sorting the leaderboard by clicking at the name of the corresponding column has been useful / a good feature.
    \item I feel that the feature of displaying all my previous submissions on the leaderboard by clicking at my name has been useful / a good feature.
    \item I feel that the inclusion of the \emph{filter} search input has been useful / a good feature.
\end{enumerate}

Possible answers for %\emph{Too much autonomy / Enough autonomy / Balanced / Guided enough / Too guided}
\textbf{Q11}: \emph{I prefer this format, just assignments and projects / I prefer evaluation via midterms}; possible answers for \textbf{Q12}:
\emph{Attacks / Defense / Equal difficulty}; all other questions had a continuous scale of answers: \emph{$1$ = Strongly disagree / $2$ = Disagree / $3$ = Neutral / $4$ = Agree / $5$ =  Strongly agree}.

\section*{Acknowledgements}
This work is supported by the National Science Foundation (NSF) under Grant DGE-2039634. We thank Yanqi Gu (UC Irvine) for assistance in the Spring 2022 course; we thank Ishana Patel (UC Irvine) for assistance in the Winter 2022 course, and we thank Hamza Errahmouni (UC Irvine) for assistance in the early stages of \emph{Maestro} development.

% Use \bibliography{yourbibfile} instead or the References section will not appear in your paper
\bibliography{aaai23}

\end{document}